\documentclass[aps,prl,reprint]{revtex4-2}
\usepackage{graphicx}% Include figure files
\usepackage[caption=false,farskip=0pt,labelfont={bf}]{subfig}
\usepackage{dcolumn}% Align table columns on decimal point
\usepackage{bm}% bold math
\usepackage{xcolor}

\begin{document}
\title{High-energy velocity tails in uniformly heated granular materials}

\author{Yangrui Chen$^1$ and Jie Zhang$^{1,2}$}
\email{jiezhang2012@sjtu.edu.cn}
\affiliation{$^1$School of Physics and Astronomy, Shanghai Jiao Tong University, Shanghai 200240, China}
\affiliation{$^2$Institute of Natural Sciences, Shanghai Jiao Tong University, Shanghai 200240, China}
		
\date{\today}

\begin{abstract}
We experimentally investigate the velocity distributions of quasi two-dimensional granular materials, which are homogeneously driven, i.e. uniformly heated, by an electromagnetic vibrator, where the translational velocity and the rotation of a single particle are Gaussian and independent.  We observe the non-Gaussian distributions of particle velocity, with the density-independent high-energy tails characterized by an exponent of $\beta=1.50\pm0.03$ for volume fractions of $0.111\le\phi\le0.832$, covering a wide range of structures and dynamics. Surprisingly, our results are in excellent agreement with the prediction of the kinetic theories of granular gas, even for an extremely high volume fraction of $\phi=0.832$ where the granular material forms a crystalline solid. Our experiment reveals that the density-independent high-energy velocity tails of $\beta=1.50$ are a fundamental property of uniformly heated granular matter.
\end{abstract}

\maketitle
\paragraph{Introduction.} 
Due to inelastic collisions\cite{RN84,RN83,RN69}, a granular gas may exhibit rich behaviors drastically different from those of a molecular gas\cite{esipov1997granular,kadanoff1999built,nagel2017experimental}, such as clustering \cite{RN61,RN62,RN90,RN110,lim2019cluster}, collective motions \cite{RN79,RN70,RN91,RN102,RN107}, phase separations \cite{RN98,RN67,RN74,RN104}, and non-Gaussian velocity distributions \cite{RN76,RN78,RN77,RN92}. The study of granular gas is important for a large variety of fields, including the glass and jamming transitions \cite{Durian-2007,Tanaka-2008, Dauchot-2016, Liu-2022}, the collective motions of active matter\cite{Chate,RN85,ramaswamy2017active}, and the applications in chemical engineering \cite{fan1999principles}, in meteorology \cite{wang2004modern}, and in astrophysics \cite{bouwman2008formation}.

One of the distinct characteristics of a granular gas is the high-energy tails of its velocity distributions. Based on the Enskog-Boltzmann equation, the kinetic theories of granular gas \cite{RN77,RN84,RN75,RN73,RN95,RN97,RN66} predict that the velocity distributions are non-Gaussian, exhibiting density-independent high-energy velocity tails $P(v)\propto exp(-c\times(v/\overline{\vert v\vert})^{\beta})$. Here $c$ is a constant related to inelasticity, and $\beta=1$ for homogeneous cooling states and $\beta=1.5$ for homogeneous driving states, in which the corresponding theories are formulated on the same mathematics \cite{RN77,RN84,RN75,RN73,RN95,RN97,RN66}. The predicted values of $\beta$ have been successfully reproduced in the state-of-the-art numerical works \cite{Zon,brey1999high,brey1999direct,puglisi1999kinetic,nie2002dynamics,poschel2006impact}

It is a great challenge to test the theories \cite{RN77,RN84,RN75,RN73,RN95,RN97,RN66} in experiments. The values of $\beta=1.0$ and $\beta=1.5$ have been reported in a number of experiments \cite{RN101,RN99,RN71,RN68,RN85,RN81,RN88, RN111,RN103}, which, however, are not exactly in line with the theoretical setup and predictions \cite{RN77,RN84,RN75,RN73,RN95,RN97,RN66}. In the three-dimensional (3D) systems driven by a uniform magnetic field along the gravitational direction \cite{RN81,RN88}, the tails of the distributions of horizontal velocity components are density-independent with $\beta=1$ instead of $1.5$, which is attributed to the anisotropy of driving and gravity and is similarly observed in \cite{RN103}. In the quasi two-dimensional (2D) vertical systems subject to the boundary driving, there is spatial heterogeneity along the vertical direction\cite{RN68,RN111}. Consequently, the results are analyzed for particles around the mid-height of the systems\cite{RN68,RN111}. In the quasi 2D horizontal layers of beads subject to homogeneous vertical shaking\cite{RN110,RN111,RN71}, different values of $\beta$ are reported in similar setups depending on the third dimension of a system, suggesting an important role of 3D effects. Interestingly, the $\beta\approx1.5$ is observed in a plasma-like dilute quasi 2D horizontal layer of beads with long-range electrostatic interactions\cite{aranson2002velocity}, where the value of $\beta$ may also depend on the driving frequency as further revealed in the numerical simulations \cite{aranson2002velocity}. In a 2D layer of homogeneously driven \emph{Vibrot} particles \cite{RN85}, the velocity distributions exhibit high-energy tails of $\beta \approx 1.50$ for volume fractions of 0.47 and 0.6. However, all \emph{Vibrot} particles rotate persistently along a single direction, and behave more like an active matter.

In this letter, we design the vibration-driven Brownian particles(\emph{VBP}) to investigate the velocity distributions of the homogeneously driven quasi 2D granular systems from the dilute gases to the ultra dense limit of granular crystals, covering a broad range of volume fractions of $0.111\le \phi\le 0.832$, in which the volume fraction $\phi$ refers to the ratio between the area covered by all particles and the total area of a 2D system. We show that the non-Gaussian velocity distributions exhibit robust density-independent high-energy tails with $\beta= 1.5\pm0.03$ for all volume fractions. Our results are in excellent agreement with the predictions of the kinetic theory of granular gas. This surprising agreement suggests that the predictions of the kinetic theory of granular gas are very general, regardless of the detailed structures and dynamics as well as the overall volume fraction.

\begin{figure}[htpb]
\centering
\includegraphics[width=8.6cm]{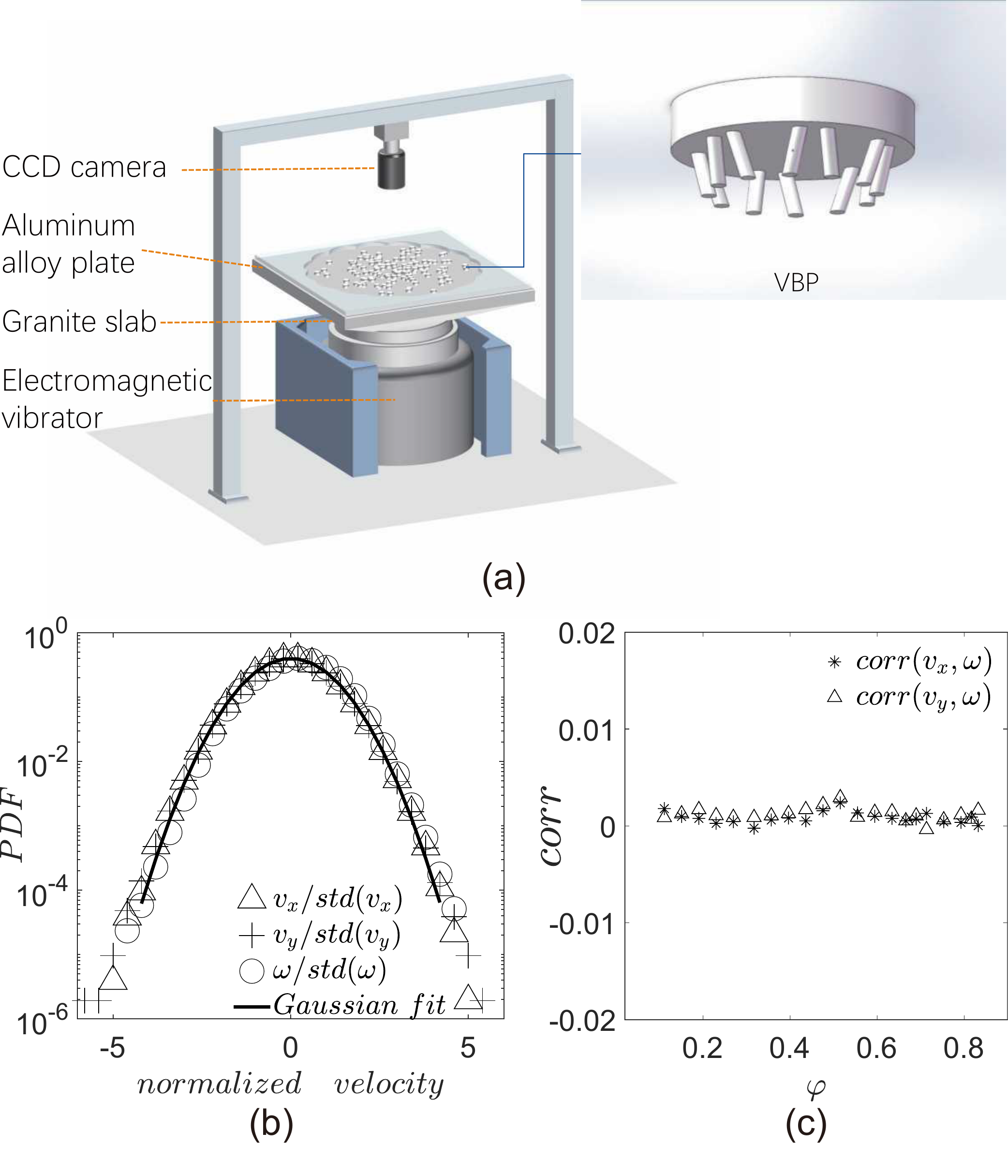}
\caption{\label{fig1} (a)Schematic of the experimental setup. The inset shows that \emph{VBP} is a disk with 12 alternately inclined supporting legs. The diameter of the disk is $D=16mm$ and the thickness is $3mm$. The legs with a height of $3mm$ are inclined inward by $18.4^\circ$ and alternately deviated from the mid-axis plane by $\pm 38.5^\circ$. We use a black marker to draw a line on the surface of each disk to identify its orientation. (b) The normalized velocity distributions of the translational components $v_x$, $v_y$ and the rotational component $\omega$ of a single \emph{VBP}. (c) The Pearson correlation coefficients between $v_x$ and $\omega$ and between $v_y$ and $\omega$, showing the translational and rotational motions of the single \emph{VBP} are independent.}
\end{figure}

\paragraph{Experimental setup.} Our experimental system is a horizontal layer of quasi 2D mono-disperse granular materials driven homogeneously by an electromagnetic vibrator as shown in Fig.~\ref{fig1}(a). The electromagnetic vibrator drives vertically a whole piece of granite with a top flat surface ($60cm \times 60cm$) of a weight of $60Kg$. The rigidity of granite effectively suppresses potential standing waves on the top flat surface. A flat aluminum alloy plate ($60cm \times 60cm \times 1cm$) is mounted onto the granite by eight strong F-clamps. An acrylic sheet is cut into a flower-shaped lateral boundary as used in Refs.\cite{RN70,RN223} to re-inject creeping particles into the interior to suppress creep particle motions along the boundary\cite{RN224}.
The electromagnetic vibrator provides sinusoidal vibration with the frequency $f=100Hz$ and the maximum acceleration $a=3g$, where $g$ is the gravitational acceleration. The amplitude $A\equiv a/(2\pi f)^2= 0.074mm$, and thus the particle displacement in the vertical direction is negligible compared to the horizontal displacement. We carefully control the vibrator to ensure that the granular system has no obvious gravitational drift within six hours of continuous vibration. \par
In each experiment, we first place a given number of \emph{VBP's} randomly and uniformly on the base plate and apply vibration for two hours to obtain an initial state. We then use a Basler CCD camera (acA2040-180kc) above to capture particle configurations and to track particle motions at 40 frames/second for one hour. We repeat each experimental run three times to take an ensemble average for an array of volume fractions $\phi$ ranging from 0.111 to 0.832 in an equal interval of 0.040. Finally, we identify and track the positions and orientations of particles using an image processing software. We discard the particles within three layers next to the boundary to eliminate potential boundary effects.

\paragraph{Single VBP.} The velocity distribution of a flat-bottomed disk on a smooth vibrating surface is non-Gaussian\cite{RN99}. Inspired by \emph{Vibrot} particles introduced by Altshuler \cite{RN65} and Scholz \cite{RN80} et al, we design a novel vibration-driven Brownian particle (\emph{VBP}) shown in the inset of Fig.~\ref{fig1}(a), whose legs are inclined inward to prevent stumble with alternately tilted angles between two neighbouring legs to effectively randomize the motion of the particle with respect to the flat surface. Before a leg bounces onto the vibrating plate, a single \emph{VBP} performs a quasi-ballistic motion with an average time $\tau_{b} \approx 38ms$, which is slightly longer than the sampling interval $\tau_{si} = 25ms$. Altogether, we have a maximum of 1100 \emph{VBP's} which are 3D printed with CBY-01 resin. 

\begin{figure}[htpb]
\centering
{\includegraphics[width=8.6cm]{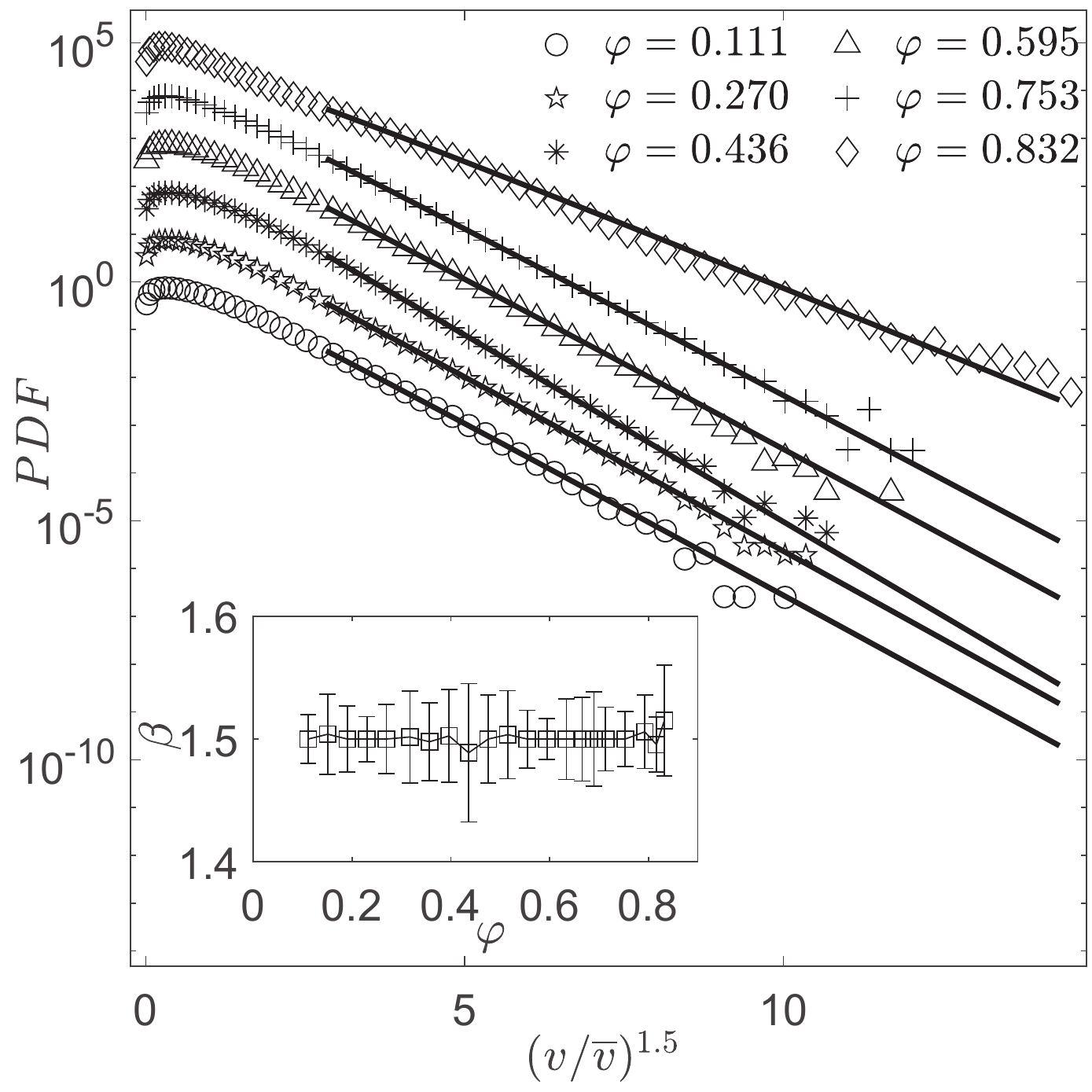}}
\caption{\label{fig2} Velocity distributions. The horizontal axis represents the normalized velocity $(v/\overline{\vert v\vert})^{1.5}$ with $v=\sqrt{(v^2_x+v^2_y)}$, where $\overline{\vert v\vert}$ is the characteristic velocity corresponding to the ensemble mean of the reduced kinetic energy $\langle \frac{1}{2}v^2 \rangle$. Here $\langle \rangle$ means average over both space and time. The results in this figure are averaged over both space and time. A set of selected representative probability-density curves are consecutively shifted upwards by a factor of $10$ to enhance visibility. The solid lines are drawn as a guide to the eye, representing the results of nonlinear fitting, and the inset shows the fitting coefficients $\beta$ for all volume fractions $\phi$.}
\end{figure}

\begin{figure}[htpb]
\centering
{
\includegraphics[width=8.6cm]{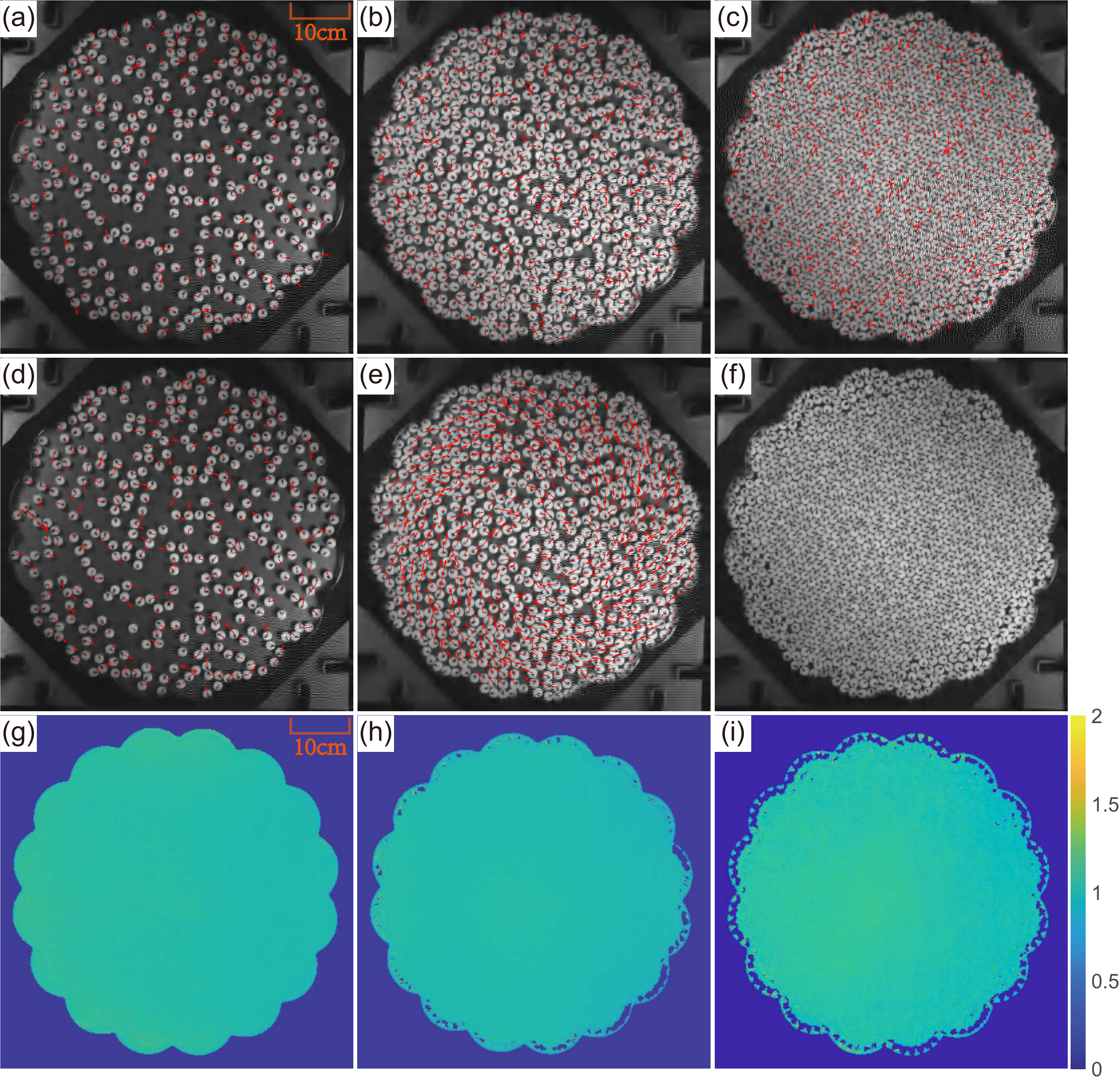}}
\caption{\label{fig3}  (a-c) Snapshots of three representative particle configurations at volume fractions $\phi=0.270,0.555,0.832$. The red arrows depict the individual particle's displacement vector during the interval of $25ms$, which corresponds to the inverse of the frame rate of video recording, i.e. $40 frames/sec$. Here, the arrow size is magnified 20 time from its original size. (d-f) The snapshots of the particles configurations are exactly the same as those in the corresponding panels of (a-c) except that the red arrows now represent the particle displacement fields during the much longer interval of 10 minutes. Here, the arrow size is reduced to 0.1 times from its original size. (g-i) The local speed map at $\phi=0.270,0.555,0.832$ showing spatially uniform driving, the color represents the time average of $\vert v_{local}\vert/v_{global}$, where the $v_{local}$ is the local velocity at a given time, i.e. the displacement over consecutive frames of time interval $25ms$, and $v_{global}$ is the square root of the spatial average of $v^2_{local}$. 
}
\end{figure}

\paragraph{Velocity distributions.} 
For a single \emph{VBP}, the distributions of rotational velocity $\omega$ and translational velocity components $v_x$ and $v_y$ are shown in Fig.~\ref{fig1}(b). All curves are normalized by their standard deviations. The results fit perfectly with the Gaussian distribution represented by the solid line with a mean value $ \vert\mu\vert < 0.1 $, which means that the motion of a single \emph{VBP} is both random and isotropic. For different \emph{VBP's}, their $\mu's$ fluctuate very near zero. In Fig.~\ref{fig1}(c), we confirm that the translation and rotation are completely independent. Thus, \emph{VBP} is a suitable model for uniformly driven quasi 2D granular materials, analogous to a uniformly heated thermal system.\par
Our main goal is to measure the velocity distributions of the uniformly heated granular materials and compare them with the predictions of kinetic theory\cite{RN77,RN84,RN75,RN73,RN95,RN97,RN66}. The velocity distributions of different volume fractions $\phi$ are shown in the main panel of Fig.~\ref{fig2}. Here we only plot the curves for a selected set of $\phi$ in the main panel of Fig.~\ref{fig2} to avoid overclouded curves.
All curves follow the similar stretched exponential fat tails on the semi-log plot, where the solid lines are guide to the eye with the fitting of the stretched exponential form $\propto exp(-c\times(v/\overline{\vert v\vert})^{\beta})$ for $v/\overline{\vert v\vert}\ge 3$. The values of the stretched exponent $\beta$ are all around $1.5$, in excellent agreement with the prediction of the kinetic theory\cite{RN77,RN84,RN75,RN73,RN95,RN97,RN66}. The values of $\beta$ within a wide range of volume fractions, i.e. $0.111\le \phi \le 0.832$, are shown in the inset of Fig.~\ref{fig2}, showing the consistent values of $\beta=1.5\pm0.03$. This fantastic agreement is rather surprising for two reason: firstly, the particle dynamics and configurations depend strongly on the volume fraction $\phi$; secondly, the dense regime where $\phi\ge0.7$ goes beyond the typical applicable regime of the kinetic theory of granular gas\cite{RN77,RN84,RN75,RN73,RN95,RN97,RN66}. It is interesting to point out that at $\phi=0.832$ where the system is a crystalline solid the distribution curve shows two distinguishable regimes for small and large $v/\overline{\vert v\vert}$ with a kink around $v/\overline{\vert v\vert}=3$. Our findings, however, disagree with the 2D simulations \cite{Zon}, where $\beta$ depends significantly on the volume fraction.

In an early experiment, a layer of stainless-steel beads is shaken vertically on the plate of an electric vibrator\cite{RN110}: when the shaking is gentle such that the beads remain quasi 2D with no particle crossing through the third dimension, the high-energy velocity tails are exponentially distributed for a low and a high volume fractions, qualitatively different from our results. We are not sure the exact cause of such exponential distributions, but we suspect that the velocity distributions of a single stainless steel bead may not be Gaussian, similar to a single flat-bottomed disk\cite{RN99}.
In a similar experiment\cite{RN71}, when the shaking is sufficiently strong such that the driven stainless-steel beads are quasi 3D, with all beads confined within the five-particle-diameter space vertically by a top glass cover, the distributions of the horizontal velocity components show high-energy tails of $\beta\approx1.5$\cite{RN71}. It is clear that the 3D effect is important for the $\beta\approx1.5$ results\cite{RN71}: when the shaking strength is tuned down to recover a quasi 2D system, the high-energy tails of velocity distributions become exponentially distributed, in agreement with the results seen in \cite{RN110}.
In another experiment of a similar geometry \cite{RN111}, $ZnO_2$ beads are confined  between a rough top cover and a flat bottom plate with an average thickness of 1.8 particle diameters within a cylindrical cell. When the cell is subject to vertical vibrations, the horizontal velocities also show  high-energy tails of $\beta\approx1.5$ \cite{RN111}. However, it is unclear how important the 3D effect is and whether the distribution of a single particle is itself non-Gaussian in \cite{RN111}. 

The results in Fig.~\ref{fig2} are nontrivial as can be seen from the qualitatively different dynamics at different volume fractions $\phi$. Figures.~\ref{fig3}(a-c) show the typical particle configurations along with the respective instantaneous velocity fields, i.e. the displacement field over the shortest time interval $25ms$ for three different volume fractions $\phi$. Here the arrows are magnified 20 times for better visibility. For all three $\phi's$, the instantaneous velocity fields are spatially random despite that at different $\phi's$, the particle configurations are quite different. At the relatively low volume fraction $\phi=0.270$, the particle configurations are more heterogeneous and sparse with chain-like structures mingled with empty space. At the higher volume fraction $\phi=0.555$, the particle configurations are more uniform in space. At $\phi=0.832$, the particles fill up the whole space in crystalline structures except near the boundary and become a solid. Figures~\ref{fig3}(d-f) show the corresponding displacement fields at much longer interval of $10$ minutes. At $\phi=0.270$ the displacement field is still spatially random as shown in panel (d), similar to the one in panel (a). However, at $\phi=0.555$ the displacement field shows a collective swarming motion in panel (e), similar to those observed in the active matter system \cite{ramaswamy2017active}. At $\phi=0.832$ the long-interval displacement field is negligibly small, in consistent with the crystalline structures. It is remarkable that with the increase of $\phi$ the particle configurations and dynamics change drastically the velocity distributions nonetheless show robust density independent high-energy tails of the same exponent $\beta=1.5\pm0.03$. 

In order to verify the spatially uniform energy input, we first coarse-grain the bottom surface with the grid size of $0.1D$, where $D=16mm$ refers to the diameter of a \emph{VBP}, and then draw the color map of the ensemble means of $\vert v_{local}\vert/v_{global}$ in Figs.~\ref{fig3}(d-f). Here $\vert v_{local}\vert$ is the local velocity magnitude at a given time, i.e. the magnitude of displacement between two consecutive frames of the time interval $25ms$, and $v_{global}$ is the square root of the spatial average of $v^2_{local}$. Except for the outermost three particle layers, the fluctuations of $\vert v_{local}\vert/v_{global}$ are less than $1\%$. We change the coarse-grain scale from $0.1D$ to $0.5D$, and the results are consistent. The similar uniform heating is also observed in the 2D homogeneously driven \emph{Vibrot} particles \cite{RN85}. 
In an early experiment, where a quasi 2D vertical cell filled with a small fraction of stainless-steel beads is subject to vertical vibrations at the bottom wall, the distribution of the horizontal velocities is given by $P(v)\propto exp(-c\times|v|^\beta)$, with $\beta$ around 1.5  \cite{RN68}. However, the boundary-driven systems typically suffer from strong spatial inhomogeneity, e.g. for the density and velocities along the vertical direction under normal gravity \cite{RN68} or micro gravity \cite{RN111}, and the results are thus only analyzed within a horizontal strip centered around the mid height of the cell \cite{RN68,RN111}.
Thus, our system is indeed subject to a spatially uniform energy input, which we believe is the key to our successful observation of the high-energy velocity tails with the predicted exponent of $\beta=1.5\pm0.03$ for all volume fractions.

The kinetic theories \cite{RN77,RN84,RN75,RN73,RN95,RN97,RN66} that lead to the prediction of the exponent of $\beta=1.5$ contain two essential ingredients as follows. Firstly, despite that kinetic energy is not conserved, momentum, or equivalently velocity, is still conserved during collisions; each collision tries to preserve the direction and magnitude of the incoming particle's velocity if its speed is much larger than that of its collision particle. Secondly, there exists a uniform heating, which is equivalent to the random walk of each particle in its velocity subspace. In the case of the homogeneous cooling of granular gas in the absence of uniform heating, the kinetic theories predict an exponential distribution for the high-energy velocity tail \cite{RN77,RN84,RN75,RN73,RN95,RN97,RN66}. This exponential tail can be understood more intuitively if we make an analogy between the momentum conservation during the collision in granular gas particles and the stress (or essentially force) conservation in jammed granular materials, where the most probable large contact force distributions are exponential \cite{liu1995force,henkes2007entropy}. In the presence of the uniform heating \cite{RN77,RN84,RN75,RN73,RN95,RN97,RN66}, the high-energy tails of velocity distributions become a stretched exponential of an exponent of $\beta=1.5$ considering the additional contribution of the random walk of each particle in its velocity subspace.

\paragraph{Conclusion} - We have built a novel system by using the vibration-driven Brownian particles to investigate the velocity distributions of the homogeneously driven quasi 2D granular materials, subject to a uniform energy input in analogous to the uniformly heated thermal systems. Our system provides the consistent results across the wide range of volume fractions of $0.111\le\phi\le 0.832$, which convincingly support that the velocity distributions exhibit the density-independent high-energy tails $\propto exp(-c\times(v/\overline{\vert v\vert})^{\beta})$, where $\beta=1.5\pm0.03$ regardless of the structures of the particles configurations and detailed dynamical behaviors. Our results agree excellently with the theoretical prediction of granular gas even for granular crystalline solids, suggesting that the density-independent high-energy tails are the fundamental characteristics of the uniformly heated granular materials.  

\begin{acknowledgments}
Y.C. and J.Z. acknowledge the support from the NSFC (No. 11974238) and from the Innovation Program of Shanghai Municipal Education Commission under No. 2021-01-07-00-02-E00138. Y.C. and J.Z. also acknowledge the support from the Student Innovation Center of Shanghai Jiao Tong University.
\end{acknowledgments}

\bibliographystyle{apsrev4-2} % Tell bibtex which bibliography style to use
\bibliography{ref} % Tell bibtex which .bib file to use (this one is some example file in TexLive's file tree)

\end{document}